\documentclass[12pt,noshowpacs,nofootinbib,notitlepage,amsmath]{revtex4-1}
\pdfoutput=1
\usepackage{setspace}
\allowdisplaybreaks
\usepackage{graphicx,color,amssymb}
\usepackage[colorlinks=true,citecolor=blue,linkcolor=blue,urlcolor=blue]{hyperref}

\def\Mp{M_{\mathrm{Pl}}}

\begin{document} 
\title{\color{blue}\Large Quadratic gravity: from weak to strong}
\author{Bob Holdom}
\email{bob.holdom@utoronto.ca}
\author{Jing Ren}
\email{jren@physics.utoronto.ca}
\affiliation{Department of Physics, University of Toronto, Toronto, Ontario, Canada  M5S1A7}
\begin{abstract}
More than three decades ago quadratic gravity was found to present a perturbative, renormalizable and asymptotically free theory of quantum gravity.
Unfortunately the theory appeared to have problems with a spin-2 ghost.
In this essay we revisit quadratic gravity in a different light by considering the case that the asymptotically free interaction flows to a strongly interacting regime. This occurs when the coefficient of the Einstein-Hilbert term is smaller than the scale $\Lambda_{\mathrm{QG}}$ where the quadratic couplings grow strong. Here QCD provides some useful insights.
By pushing the analogy with QCD, we conjecture that the nonperturbative effects can remove the naive spin-2 ghost and lead to the emergence of general relativity in the IR.

\vspace{1em}
\begin{center}
Essay awarded fourth prize in the Gravity Research Foundation 2016 essay competition.
\end{center}
\end{abstract}
\maketitle
\newpage

As a \emph{perturbative}, \emph{renormalizable} and \emph{asymptotically free} theory of quantum gravity, quadratic gravity was actively studied in the 1980's. But due to the difficulties that we shall discuss and the development of a new promising perturbative approach to quantum gravity, namely string theory, work on quadratic gravity quickly diminished. Now let us consider it once again. In four dimensions the action can be organized as follows
\begin{eqnarray}\label{eq:quadratic}
S_{\mathrm{QG}}=\int d^4x\,\sqrt{-g}\left(\frac{1}{2}M^2R-\frac{1}{2 f_2^2}C_{\mu\nu\alpha\beta}C^{\mu\nu\alpha\beta}+\frac{1}{3 f_0^2}R^2\right).
\end{eqnarray}
With only two independent quadratic terms,
the action is characterized by two dimensionless couplings and one mass scale.
As we develop a different perspective on this theory below, we shall see that the mass scale $M$ need \emph{not} be identified as the Planck mass $\Mp$.

The renormalizability of quadratic gravity (\ref{eq:quadratic}) \cite{Stelle:1976gc} comes from the dominance of quadratic terms in the UV, which contribute a $1/k^4$ dependence in the propagator of the metric fluctuation $h_{\mu\nu}$. Renormalizability is also related to the fact that besides the trace anomaly, the classical scale invariance is only softly broken by $M$. The trace anomaly corresponds to the logarithmic running of the two dimensionless couplings, which are found to be asymptotically free \cite{Fradkin:1981iu}. If the couplings are weak on the mass scale, $\sim \left|f_i M\right|$, of the massive gravitational modes,
then the theory remains perturbative. At energies below this mass scale the theory is described by general relativity (GR) with $M$ identified as $\Mp$.

In this standard view the presence of higher derivatives in the theory leads to a problem.
Inspecting the propagator of $h_{\mu\nu}$ on a flat background, the spin-2 sector exhibits a normal massless pole as well as a massive pole with negative residue. This massive ghost pole can be interpreted either as a state of negative norm or a state of negative energy. The former has difficulties to reconcile with the probability interpretation and unitarity, while the later implies a vacuum decay into negative energy ghosts and normal particles with an infinite phase space.

This perturbative analysis is correct when $M$ is sufficiently large and the theory remains perturbative. But if $M$ is sufficiently small, another mass scale appears, $\Lambda_{\mathrm{QG}}$, the scale where the couplings $f_i$ grow strong. Then the poles appearing in the perturbative propagator fall into the nonperturbative region. The previous arguments implicitly assume that the true physical spectrum is reflected by the tree-level propagator, and this need no longer be true.
The question is whether the theory can evade the ghost problem in the strong coupling regime when $M\lesssim\Lambda_{\mathrm{QG}}$.

For insight into this question, it is instructive to consider another renormalizable and asymptotically free theory, quantum chromodynamics (QCD). On one hand we are used to a physical spectrum that bears no relation to the perturbative degrees of freedom. In particular
the gluon is not in the physical spectrum. On the other hand we use the tree-level gluon propagator in perturbative QCD to calculate hard processes involving high virtuality, which are thus insensitive to the IR physics. These two perspectives are brought together by considering the full nonperturbative gluon propagator, where the existence and the position of the poles are \emph{gauge independent} concepts.
Various approaches to nonperturbative QCD indicate that the full gluon propagator is suppressed in the IR, and a mass gap develops without a massive pole and without breaking the gauge symmetries.

\begin{figure}[!h]
  \centering%
{ \includegraphics[width=7.4cm]{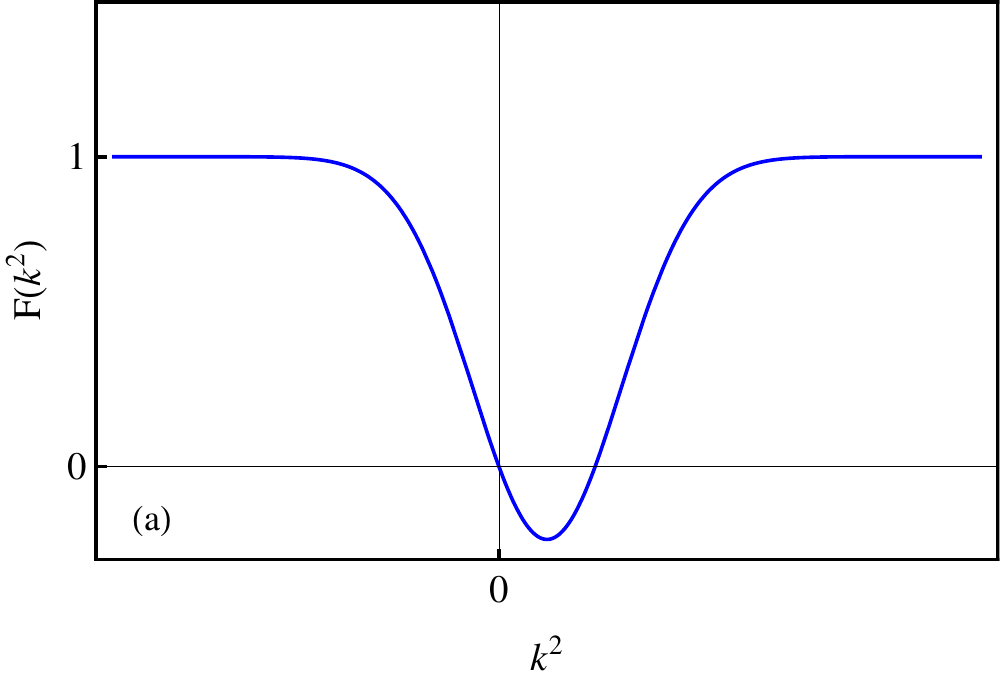}}\quad\quad
{ \includegraphics[width=7.4cm]{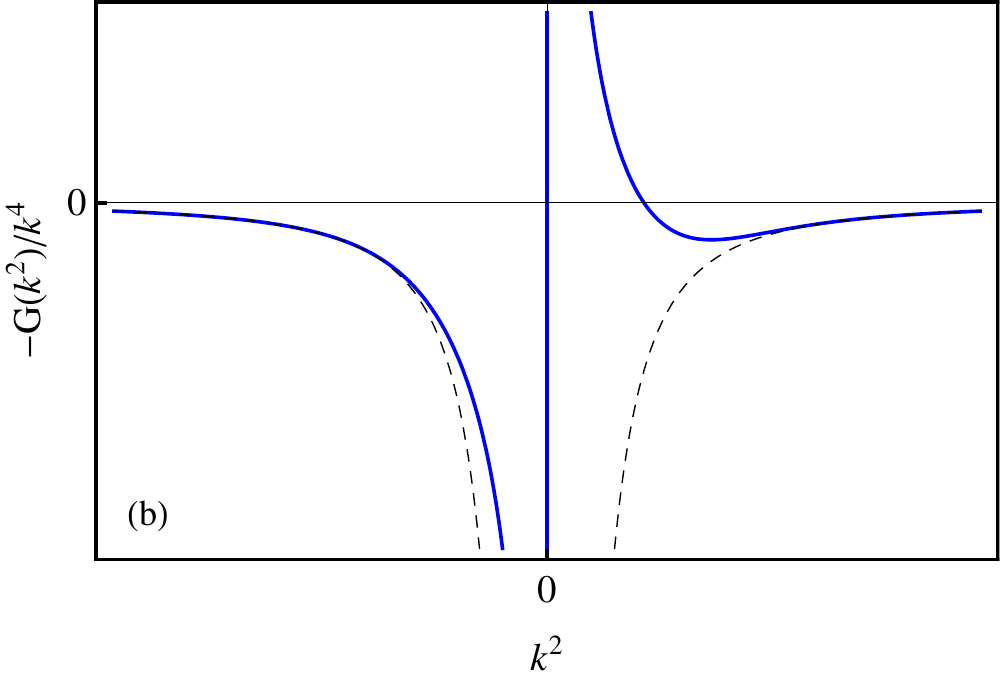}}\\
{ \includegraphics[width=7.4cm]{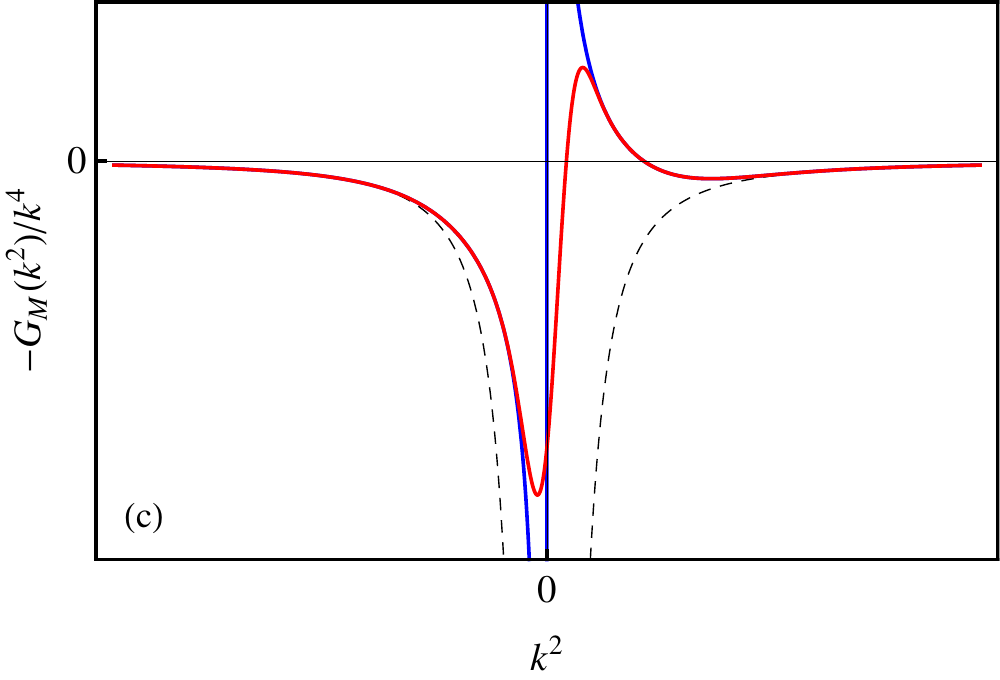}}
\caption{\label{fig:Propagator} (a) The multiplicative factor $F(k^2)$ in the full gluon propagator as suggested by the lattice studies in Landau gauge. (b) The nonperturbative graviton propagator $-G(k^2)/k^4$ for $M=0$ and a flat background (blue solid line). The black dash line denotes the tree-level propagator. (c) The nonperturbative graviton propagator $-G_M(k^2)/k^4$ for nonvanishing $M$ (red solid line).
}
\end{figure}

Recent progress in lattice studies that implement gauge fixing has led to a better understanding of this phenomenon.
We could parameterize the full gluon propagator as a real scalar factor $F(k^2)/k^2$ times a tensor factor and a perturbative correction factor. $F(k^2)$ encodes the nonperturbative modification and is nontrivial only in the IR. The most reliable evidence
comes from the lattice studies in Landau gauge~\cite{LGlattice1}. They indicate that $F(k^2)/k^2$ approaches a negative constant in the zero Euclidean momentum limit. With no massive pole, the general shape of $F(k^2)$ is plotted in Fig.~\ref{fig:Propagator}(a). The fact that $F(k^2)$ has a zero at $k^2=0$ is a gauge independent statement, while the order of this zero may be gauge dependent.

Could a similar story hold for quadratic gravity?
The analogy is most straightforward for the $M=0$ case in (\ref{eq:quadratic}) and for a flat background, where the theory has only one scale, $\Lambda_{QG}$, as in massless QCD. Similarly we parameterize the graviton propagator as $-G(k^2)/k^4$ times tensor and perturbative correction factors. Here the tree-level propagator $-1/k^4$ is the degenerate limit of the massless pole and the massive ghost pole as $M\to 0$.
If we wish to consider the possibility that nonperturbative corrections have a similar effect here as they do in QCD, then we can take $G(k^2)$ to have the same form of $F(k^2)$ as in Fig.~\ref{fig:Propagator}(a). As shown by the blue solid line in Fig.~\ref{fig:Propagator}(b) the $-1/k^4$ behavior has been softened to $1/k^2$ with positive residue, which is the normal propagator for a spin-2 massless graviton.

If general covariance is still maintained by the strong
dynamics, just as gauge invariance is preserved in QCD, the leading order effective theory to describe the massless graviton in low energy would be GR, with the Planck mass identified as
\begin{eqnarray}
\Mp^2=-1/G'(0)\sim\Lambda_{\mathrm{QG}}^2\,.
\end{eqnarray}
Similar to the chiral Lagrangian, there would be a derivative expansion of the curvature tensors with order one coefficients in Planck units.
As the fundamental theory is diffeomorphism invariant, this emergence of a spin-2 massless graviton is not in contradiction with the Weinberg-Witten theorem \cite{WW}.
So we arrive at a picture of the theory that is weakly interacting in both the UV and IR limits, with only an intermediate region that is strong.

What happens for the theory with a nonvanishing $M\lesssim \Lambda_{\mathrm{QG}}$?
We have mentioned that $M$ falling below a certain value corresponds to the naive perturbative masses dropping below $\Lambda_{\mathrm{QG}}$.
We propose that $M$ gives control over a \emph{quantum phase transition} between the weak and strong phases, where the latter has a distinctly different physical spectrum. In terms of the full graviton propagator, the impact of $M$ could be encoded in a new form factor $G_M(k^2)$. One possibility for the full propagator is illustrated by the red solid line in Fig.\ref{fig:Propagator}(c).
In the intermediate window $M^2\ll k^2\ll\Mp^2$, the propagator behaves approximately as $1/k^2$. In the deep IR where $k^2<M^2$, the additional suppression due to $M$ removes the would-be massless pole. The \emph{mass gap}, although characterized by a scale different from $\Lambda_{\mathrm{QG}}$, is analogous to that of the gluon with no massive pole and intact gauge invariance. $G_M(k^2)$ has a second order zero, and the result is physically different from the previous case of a first order zero. The graviton would only propagate as a virtual particle and clearly $M$ would have to be extremely small for this to be realistic.

Finally, should we still wonder how a consistent quantum theory can emerge from a problematic tree-level action? Here we use the path integral formulation as a nonperturbative definition of the quantum theory. For a gauge theory a nontrivial measure is constructed to uniformly sample gauge orbits in the configuration space of the path integral. This procedure effectively alters the action and a similar construction applies to both QCD and gravity. In QCD it is known that the measure brings in effects connected with the nontrivial structure of the gauge configuration space, i.e. Gribov copies. These nonlocal and nonperturbative effects
are intimately related to the absence of a gluon pole. This motivates the study of Gribov copies for gravity. Similar to the initial indication of Gribov copies in QCD~\cite{Gribov:1977wm} we have found infinitesimally close pairs of copies in a covariant gauge~\cite{Holdom:2015kbf}. Thus from the path integral definitions we find some evidence for the fundamental similarity between quadratic gravity and QCD.

Such a picture of quadratic gravity has interesting physical implications. Taking the $M=0$ case in particular, the physics at low energy or curvature is described by GR as an effective theory, and the semiclassical description breaks down around $\Mp$.
But in contrast to the standard picture the effects of strong gravity only occur in a limited range. In the far UV the couplings become arbitrarily small and the theory could be fundamentally defined around flat spacetime.
The asymptotically free nature of gravity implies that the matter sector to which it couples should also be UV complete.

Quadratic gravity has new static spherically symmetric vacuum solutions that resemble black holes while remaining \emph{horizonless}~\cite{QGBH}. In the small curvature exterior region, the solution is Schwarzschild-like but with exponentially small Yukawa-like corrections from the higher order terms. When the would-be horizon is approached, the Yukawa terms quickly drive the curvatures to be Planck-sized and a strongly interacting shell appears instead of a horizon. At small radius the curvature of the vacuum solution may continue to increase so that the physics is largely governed by the classical quadratic action. The first direct detection of gravitational waves from a compact-binary coalescence by aLIGO opens up the era of testing nonlinear gravity~\cite{aLIGO}. In this new window the horizonless object may provide an opportunity for the effects of large curvatures and strong quantum gravity to be revealed.

\linespread{1}

\begin{thebibliography}{99}


\bibitem{Stelle:1976gc}
  K.~S.~Stelle,
  Phys.\ Rev.\ D {\bf 16}, 953 (1977).
  B.~L.~Voronov and I.~V.~Tyutin,
  Yad.\ Fiz.\  {\bf 39}, 998 (1984).


\bibitem{Fradkin:1981iu}
  E.~S.~Fradkin and A.~A.~Tseytlin,
  Nucl.\ Phys.\ B {\bf 201}, 469 (1982).
  I.~G.~Avramidi and A.~O.~Barvinsky,
  Phys.\ Lett.\ B {\bf 159}, 269 (1985).



\bibitem{LGlattice1}
  A.~Cucchieri and T.~Mendes,
  PoS LAT {\bf 2007}, 297 (2007)
  [arXiv:0710.0412 [hep-lat]].
  I.~L.~Bogolubsky, E.~M.~Ilgenfritz, M.~Muller-Preussker and A.~Sternbeck,
  PoS LAT {\bf 2007}, 290 (2007)
  [arXiv:0710.1968 [hep-lat]].


\bibitem{WW}
  S.~Weinberg and E.~Witten,
  Phys.\ Lett.\ B {\bf 96}, 59 (1980).



\bibitem{Gribov:1977wm}
  V.~N.~Gribov,
  Nucl.\ Phys.\ B {\bf 139}, 1 (1978).
  I.~M.~Singer,
  Commun.\ Math.\ Phys.\  {\bf 60}, 7 (1978).



\bibitem{Holdom:2015kbf}
  B.~Holdom and J.~Ren,
  arXiv:1512.05305 [hep-th].

\bibitem{QGBH}
  B.~Holdom,
  Phys.\ Rev.\ D {\bf 66}, 084010 (2002)
  [hep-th/0206219].
  H.~L¨¹, A.~Perkins, C.~N.~Pope and K.~S.~Stelle,
  arXiv:1508.00010 [hep-th].


\bibitem{aLIGO}
  B.~P.~Abbott {\it et al.} [LIGO Scientific and Virgo Collaborations],
  Phys.\ Rev.\ Lett.\  {\bf 116}, no. 6, 061102 (2016)
  [arXiv:1602.03837 [gr-qc]].


\end{thebibliography}
\end{document}